\documentclass[twocolumn,amsmath,amssymb,aps]{revtex4}

\usepackage{graphicx}
\usepackage{dcolumn}
\usepackage{bm}
\usepackage{textcomp}
\usepackage{url}
\DeclareGraphicsExtensions{.jpg,.eps,.pnm}

\begin{document}
\title{A vertically-coupled liquid-crystal long-range plasmonic optical switch} \vspace{0.15cm}
\author{Dimitrios C. Zografopoulos$^*$ and Romeo Beccherelli}

\affiliation{\vspace{0.15cm}  Consiglio Nazionale delle Ricerche, Istituto per la Microelettronica e Microsistemi (CNR-IMM), Roma 00133, Italy. $^*$E-mail: dimitrios.zografopoulos@artov.imm.cnr.it  \vspace{0.15cm}}
\date{\today}

\begin{abstract}
An optical switch based on liquid-crystal tunable long-range metal stripe waveguides is proposed and theoretically investigated. A nematic liquid crystal layer placed between a vertical configuration consisting of two gold stripes is shown to allow for the extensive electro-optic tuning of the coupler's waveguiding characteristics. Rigorous liquid-crystal switching studies are coupled with the investigation of the optical properties of the proposed plasmonic structure, taking into account different excitation conditions and the impact of LC-scattering losses. A directional coupler optical switch is demonstrated, which combines low power consumption, low cross-talk, short coupling lengths, along with sufficiently reduced insertion losses.
\end{abstract}

\maketitle

Intense scientific research effort has being recently directed towards the field of guided-wave plasmonics, which is concerned with the manipulation and routing of strongly localized surface plasmon polaritons (SPPs), i.e., light waves coupled to oscillations of a metal's free electrons that propagate along metal/dielectric interfaces \cite{Gramotnev2010}. Various plasmonic structures have been thus far proposed as the essential elements of an integrated photonics platform for broadband light routing and optical signal processing. Among these, long-range (LR) waveguides, composed of thin metal stripes, have been shown to offer low propagation losses, mode matching to single-mode fibers, planar processing fabrication and the capability of controlling the optical signal by direct addressing the metal waveguides \cite{Berini2009}. Such capability allowed for the demonstration of numerous tunable devices indispensable in real optical signal processing architectures, such as modulators and switches, which are based on the thermo-optic control of the background polymer via current injection through the metal waveguide \cite{Nikolajsen2004}, \cite{Boltasseva2006}, \cite{Charbonneau2006}.

Apart from the exploitation of the thermo-optic effect, a traditional alternative route aiming at dynamically tunable photonic devices involves the electro-optic control of nematic liquid crystals (NLCs), inherently anisotropic materials whose properties are highly responsive to the application of external fields. Already established in the design and fabrication of dielectric-based photonic guided-wave components \cite{Zografopoulos2012b}, NLCs are furthermore currently being explored as a tuning mechanism of the properties of plasmonic structures \cite{Abdulhalim2012}.  Functional LC-based plasmonic components, such as variable attenuators, phase-shifters and switches have already been theoretically demonstrated in a variety of waveguide platforms \cite{Zografopoulos2012c}, \cite{Zografopoulos2012d}, \cite{Tasolamprou2011a}. Based on capacitive operation, LC-plasmonic devices are not only free from current injection issues, such as electromigration or thermal diffusion crosstalk, but, more importantly, they allow for extremely low-power consumption, several orders of magnitude lower than their thermo-optic counterparts \cite{Pfeifle2012}.

%%%%%%%%%%%%%%%%%%%%%%%%%% FIGURE %%%%%%%%%%%%%%%%%%%%%%%%%%%%%
\begin{figure}[b]
\centering
\includegraphics[width=8.5cm]{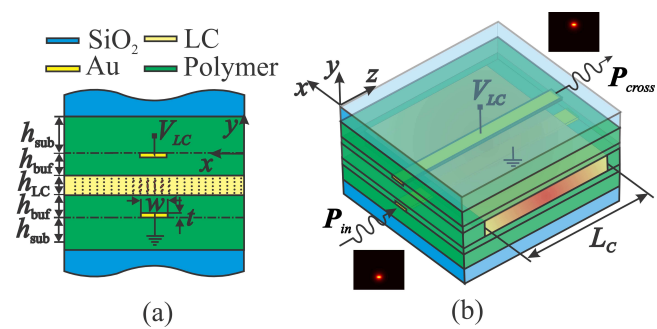}
\caption{(a) Cross-section and (b) perspective view of the proposed vertical liquid-crystal plasmonic switch and definition of material and structural parameters.}\label{figure1}
\end{figure}
%%%%%%%%%%%%%%%%%%%%%%%%%%%%%%%%%%%%%%%%%%%%%%%%%%%%%%%%%%%%%%%%%

In this work we propose a LC-tunable LR-SPP optical switch, based on the vertical-coupling configuration shown in the schematic layout of Figure \ref{figure1}. Such structures provide a route towards the development of three-dimensional multilayered architectures for photonic integrated circuits and printed circuit boards that offer more compact interaction lengths than purely co-planar configurations \cite{Won2006}, \cite{Srivastava2010}. Two identical gold stripes are embedded in a polymer background medium, benzo-cyclobutene (BCB), and they are separated by a LC cell infiltrated with the known nematic mixture E$7$. The structure is fabricated on a low-index substrate that does not affect its optical properties, in this case silica glass. The design parameters here selected involve materials typical of experimentally demonstrated LR-SPP devices and allow for the demostration of the salient features of the proposed switch. In principle, the only requirement is to meet the condition $n_o < n_p < n_e$, where $n_o$, $n_e$ are the ordinary and extraordinary LC indices and $n_p$ the polymer refractive index, for reasons that will be made clear. Geometrical parameters, as defined in Figure \ref{figure1}(a), are set to $h_{\textrm{sub}} = 15$ {\textmu}m, $h_{\textrm{buf}} = 7$ {\textmu}m, $h_{\textrm{LC}} = 3$ {\textmu}m, $w = 5$ {\textmu}m, and $t=15$ nm. A voltage difference $V_{\textrm{LC}}$ is applied via the set of two gold stripe waveguides. The LC material is characterized by ordinary and extraordinary relative permittivities $\varepsilon_o = 5.3$ and $\varepsilon_e = 18.6$, and elastic constants $K_{11}$, $K_{22}$, $K_{33}$ equal to $10.3$, $7.4$, and $16.48$ pN, respectively \cite{Stromer2006}. Relative permittivities of BCB and SiO$_2$ are equal to $2.65$ \cite{DowChemicals} and $3.91$, respectively. Material dispersion is provided via the Cauchy model for BCB \cite{DowChemicals} and E7 \cite{Li2005}, the Sellmeier model for SiO$_2$ \cite{Tatian1984}, and interpolated experimental data for Au \cite{Palik1985}. Their corresponding values at $\lambda_0=1.55$ {\textmu}m are $n_{\textrm{BCB}} = 1.535$, $n_o = 1.5024$, $n_e = 1.697$, $n_{\textrm{SiO}_2} = 1.444$, and $n_{\textrm{Au}} = 0.5 - j11.5$.

The operation of the switch relies on the electro-optic control of the LC molecular orientation. When no external voltage is applied, the LC molecules are aligned along the $z$-axis. Strong anchoring conditions are imposed at the polymer/LC interfaces via, for instance, rubbed alignment layers. In this state, the TM-polarized light supported by the long-range mode of the Au stripes senses a low-index LC layer, since $n_o<n_p$. Thus, light coupling between the two plasmonic waveguides is inhibited and the switch unconditionally operates in the BAR state, i.e., light propagates along and finally exits from the same waveguide it was excited. The application of $V_{\textrm{LC}}$ induces the switching of the LC molecules, primarily in the region between the two electrodes, where the electrostatic field is stronger, thus forming a high-tilt zone. For the current choice of materials, there exists a voltage threshold for which the effective index sensed by TM light, the only polarization supported by the metal stripes, exceeds $n_p$. Under such a condition, a voltage-induced LC-dielectric waveguide is formed in-between the two metal stripes, as schematically depicted in Figure \ref{figure1}(a). The configuration of the three single-mode waveguides, two plasmonic and the LC-dielectric one, is known to support two symmetric and one anti-symmetric supermodes whose effective indices $n_{\textrm{eff},i}$ are a function of the applied voltage. When the matching condition $n_{\textrm{eff},1}-n_{\textrm{eff},3}=n_{\textrm{eff},3}-n_{\textrm{eff},2}$ is satisfied, $n_{\textrm{eff},3}$ being the index of the anti-symmetric supermode, complete light transfer can be achieved to the CROSS state after a propagation distance equal to the coupling length $L_C = 0.5 \lambda / \Delta n_{\textrm{eff}}$, where $\Delta n_{\textrm{eff}} = n_{\textrm{eff,1}} - n_{\textrm{eff,3}}$, by properly exciting with a $1:1:2$ linear combination of the three supermodes \cite{Saitoh2008}.

%%%%%%%%%%%%%%%%%%%%%%%%%% FIGURE %%%%%%%%%%%%%%%%%%%%%%%%%%%%%
\begin{figure}[t]
\centering
\includegraphics[width=8.5cm]{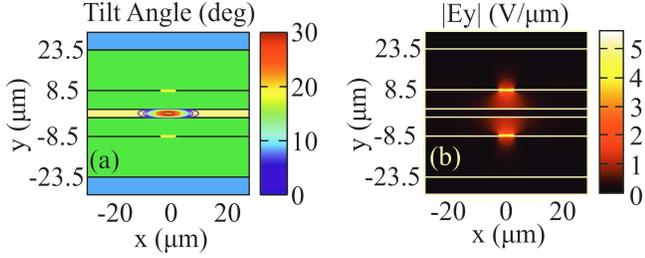}
\caption{(a) Tilt angle profile and (b) y-component of the electrostatic field for an applied voltage $V_{\textrm{LC}} = 20.58$ V, which provides a coupling length $L_C = 2.348$ mm at $1.55$ {\textmu}m.}\label{figure2}
\end{figure}
%%%%%%%%%%%%%%%%%%%%%%%%%%%%%%%%%%%%%%%%%%%%%%%%%%%%%%%%%%%%%%%%%

Based on this principle, we investigate into the performance of the proposed plasmonic switch, by considering two case-scenarios in terms of the structure excitation, namely launching the LR-SPP mode supported by the single Au stripe in the presence of a non-switched ($V_{\textrm{LC}}=0$) LC-layer, and the one supported by the single Au stripe fully embedded in BCB, which reflects the realistic experimental and fabrication conditions, as shown in Figure \ref{figure1}(b). The LC orientation profile is calculated by minimizing the total energy of the system, taking into account the coupling between the underlying electrostatic and the elastic problems, as explained in detail in \cite{Zografopoulos2012d}. This analysis provides the tilt and twist angles, which describe the deviation of the local nematic orientation from the $x-z$ plane and $z-$axis, respectively, and determine the exact form of the optical dielectric tensor. These are fed into a fully-anisotropic finite-element eigensolver \cite{Comsol}, which yields the modal indices and field distributions of the supported supermodes with respect to the applied voltage $V_\textrm{LC}$ and operation wavelength $\lambda_0$. Subsequently, the switch is excited at the bottom Au stripe by the equivalent single-waveguide LC- or BCB-superstratum LRSPP modes. The input excitation ($\mathbf{e}_i$) is projected onto the set of the voltage-dependent eigenmodes ($\mathbf{e}_m$) and is expanded to this vector-basis. This allows for the calculation of the output field distribution ($\mathbf{e}_o$) at any propagation length according to the superposition of the analytically propagated eigenmodes
\begin{equation}\label{eq0}
\mathbf{e}_o(x,y,z)=\sum_m \gamma_m \mathbf{e}_m(x,y) \exp(-j n_{\textrm{eff}(m)} k_0 z),
\end{equation}
where each eigenmode $m$ is excited with a relative weight given by the complex-valued vector overlap integral \cite{Stallein2005}
\begin{equation}\label{eq1}
\gamma_m = \dfrac{\iint_{A_{\infty}} \mathbf{e}_i \times \mathbf{h}_m^* \cdot \mathbf{\hat z} \, \textrm{d}S}{ \iint_{A_{\infty}} \mathbf{e}_m \times \mathbf{h}_m^* \cdot \mathbf{\hat z} \, \textrm{d}S}.
\end{equation}

Power transfer along the device for a given input is monitored by the output-port crosstalk XT, defined as the ratio of the guided-power $\textrm{XT} = 10 \log (P_{\parallel}/P_{\textrm{X}})$, where $P_{\parallel}$, $P_{\textrm{X}}$ is the power percentage projected to the output bar and cross-modes, respectively. Radiation modes are also included in the eigenmode-expansion analysis given that the input fields are not in general linear superpositions of the guided supermodes only. Nevertheless, it has been numerically verified that these act only as an insertion-loss mechanism, providing negligible coupling to the reference cross and bar-modes.

%%%%%%%%%%%%%%%%%%%%%%%%%% FIGURE %%%%%%%%%%%%%%%%%%%%%%%%%%%%%
\begin{figure}[t]
\centering
\includegraphics[width=8.5cm]{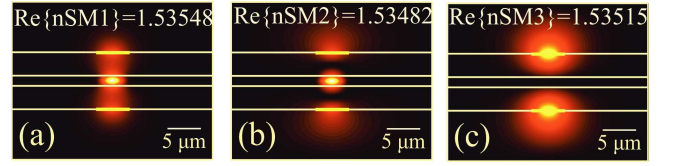}
\caption{Electric field profiles $\left| \mathbf{E} \right|$ for the three supermodes supported by the coupler at $V_{\textrm{LC}} = 20.58$ V, which satisfies at $1.55$ {\textmu}m the phase-matching condition among their indices: $n_{\textrm{eff,1}} - n_{\textrm{eff,3}} = n_{\textrm{eff,3}} - n_{\textrm{eff,2}}$.}\label{figure3}
\end{figure}
%%%%%%%%%%%%%%%%%%%%%%%%%%%%%%%%%%%%%%%%%%%%%%%%%%%%%%%%%%%%%%%%%

Figure \ref{figure2} shows the tilt angle profile and the $y-$component of the electrostatic field, for an applied voltage $V_{\textrm{LC}}=20.58$ V. The maximum tilt value is ca. $30^{\circ}$, which translates into an TM-effective index of $1.553$ at $1.55$ {\textmu}m. Twist values were found lower than $1^{\circ}$, owing to the symmetry of the structure. Given the lower permittivity of BCB, the electrostatic field is stronger in the polymer than in the LC region, as demonstrated in Figure \ref{figure2}(b).  This particular voltage value was found to satisfy the phase-matching condition among the three supermodes at the target wavelength of $1.55$ {\textmu}m, whose electric field profile is shown in Figure \ref{figure3}. The corresponding coupling length of the device is found equal to $L_C=2.348$ mm.

%%%%%%%%%%%%%%%%%%%%%%%%%% FIGURE %%%%%%%%%%%%%%%%%%%%%%%%%%%%%
\begin{figure}[t]
\centering\includegraphics[width=8.5cm]{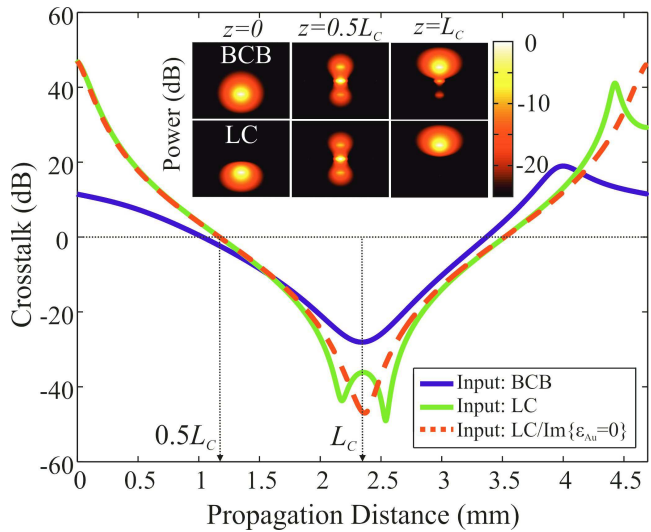}
\caption{Crosstalk evolution along propagation, when the coupler is excited by the LC- and BCB-reference LRSPP modes, at $1.55$ {\textmu}m, where $L_C=2.348$ mm. The ideal lossless case is also included for comparison. Inset shows modal power profiles at the input, output of the coupler and after a propagation distance $z=0.5 L_C$.}\label{figure4}
\end{figure}
%%%%%%%%%%%%%%%%%%%%%%%%%%%%%%%%%%%%%%%%%%%%%%%%%%%%%%%%%%%%%%%%%

Figure \ref{figure4} shows the crosstalk variation along propagation for the two excitation scenarios, namely BCB- and LC-input, at $\lambda_0=1.55$ {\textmu}m. In order to help comparison, the dashed line is also included, which corresponds to the LC-input case, where Au losses have been intentionally omitted. In that case, $\textrm{XT}(z)$ is symmetric around $L_C$ obtaining extremely low values, below $-50$ dB, at $z=L_C$. The inclusion of Au losses induces a slight detuning in the phase-matching condition, typical in lossy photonics directional couplers, which manifests as the small bump in crosstalk around $L_C$ \cite{Powelson1998}. In the BCB-input scenario, power transfer is further detuned, as the input field does not project symmetrically onto the three guided supermodes, leading to a small amount of residual optical power at the exit of the coupler, mainly in the LC layer, as demonstrated in the inset of Figure \ref{figure4}. Nevertheless, under such realistic conditions crosstalk values better than $-20$ dB are obtained at $z=L_C$. The insertion losses of the device, owing  to propagation and mode-mismatch coupling losses were calculated equal to $1.45/0.55$ dB (CROSS) and $1.62/0.82$ dB (BAR) for the BCB- and LC-input case, respectively.

%%%%%%%%%%%%%%%%%%%%%%%%%% FIGURE %%%%%%%%%%%%%%%%%%%%%%%%%%%%%
\begin{figure}[t]
\centering
\includegraphics[width=8.5cm]{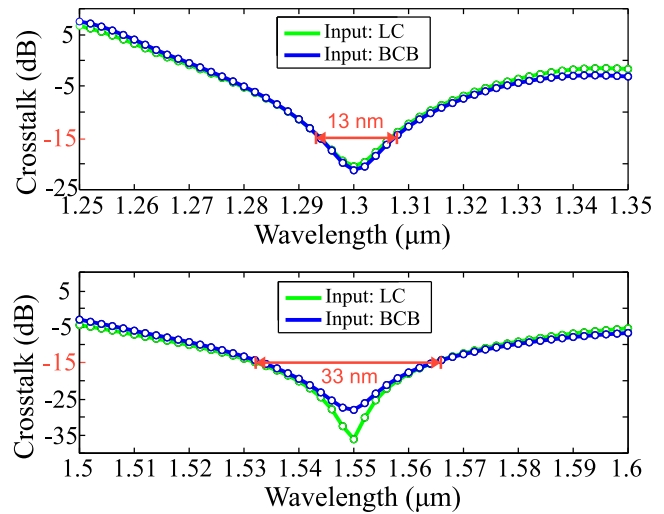}
\caption{Spectral response of the proposed LC-LRSPP switch in the $1.3$ and $1.55$ {\textmu}m telecom windows. The bandwidth providing crosstalk values better than $-15$ dB is equal to $13$ and $33$ nm, respectively. The LC voltage and coupler length for each window are fixed to those for optimized performance at the central $\lambda_0$.}\label{figure5}
\end{figure}
%%%%%%%%%%%%%%%%%%%%%%%%%%%%%%%%%%%%%%%%%%%%%%%%%%%%%%%%%%%%%%%%%

The investigation into the coupler's performance is next extended to the $1.3$ {\textmu}m telecom wavelength. The corresponding analysis yielded an optimal switching voltage equal to $V_{\textrm{LC}}^{1300} = 20.248$ V, a coupling length $L_{\textrm{C}}^{1300} = 4.02$ mm, and insertion losses equal to $2.62/2.5$ dB (CROSS) and $3.45/3.35$ dB (BAR) and for the two input scenarios. Furthermore, the broadband performance of the coupler is also investigated in $100-$nm windows centered at the two target $\lambda_0$, taking into account the material dispersion of all materials involved, with results presented in Figure \ref{figure5}. Setting a target crosstalk value of $-15$ dB, the bandwidth of the device is found equal to $33$ and $13$ nm at $1.55$ and $1.3$ {\textmu}m, respectively. The crosstalk has been calculated at a propagation distance equal at the coupling lengths of the two central $\lambda_0$. Owing to higher field confinement at lower wavelengths, minimum crosstalk values are almost identical for both excitations in the $1.3$ {\textmu}m window, with a value lower than $-20$ dB at $\lambda_0$.

\begin{figure}[t]
\centering
\includegraphics[width=8.5cm]{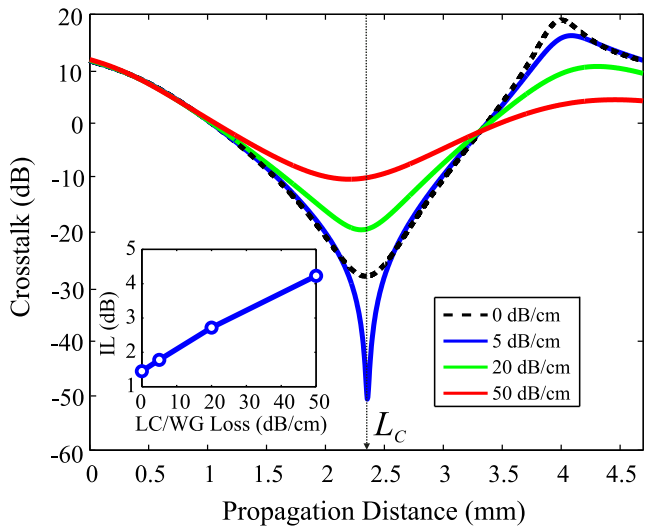}
\caption{Impact of LC-waveguide losses on the switch performance, in terms of crosstalk and insertion lossed.}\label{figure6}
\end{figure}

The analysis thus far neglected the scattering losses of the LC-dielectric waveguide that, in general, depend on the nematic material used, the geometry and the anchoring conditions. Propagation losses in LC-waveguides similar to the one here proposed have been measured typically in the range between few \cite{dAlessandro2006,Donisi2010} to tens of dB/cm \cite{Pfeifle2012}. In order to estimate the impact of LC-induced losses in the performance of the switch, we simulated the single LC-waveguide at $V_{\textrm{LC}}$ by correspondingly adding an imaginary part to the LC indices and by tuning its value so that propagation losses of the waveguide are equal to $5, 20,$ and $50$ dB/cm. Using these values, we repeated the analysis for the target wavelength of $1.55$ {\textmu}m, focusing on the realistic BCB-input scenario. Corresponding results are summarized in Figure \ref{figure6}, where the reference case of zero LC-losses is also included. Interestingly, for LC-waveguide losses of $5$ dB/cm \cite{dAlessandro2006,Donisi2010} the coupler's performance is improved compared to the zero loss case, with very low crosstalk values at a propagation distance equal to $L_C$. This can be explained by examining the related output field profile shown in the inset of Figure \ref{figure4}, where it is shown that the residual optical power lies mainly in the LC-layer. Thus, this level of losses, which is comparable to that of the single reference LR-SPP BCB-input mode ($2.45$ dB/cm), is shown capable of improving the coupling crosstalk by removing the parasitic residual power in the LC layer. No significant overall phase detuning is observed in this case. As LC-waveguide losses obtain higher values, power coupling is hindered by further detuning, the crosstalk, however, remains close to $-20$ dB, even for losses up to $20$ dB/cm. Not unexpectedly, the introduction of LC-losses also raises the overall insertion losses of the device in the CROSS state, as shown in the inset of Figure \ref{figure6}.

%%%%%%%%%%%%%%%%%%%%%%%%%% FIGURE %%%%%%%%%%%%%%%%%%%%%%%%%%%%%
%%%%%%%%%%%%%%%%%%%%%%%%%%%%%%%%%%%%%%%%%%%%%%%%%%%%%%%%%%%%%%%%%

In brief, we have presented the design and rigorous analysis of a novel long-range plasmonic switch, which shows both low crosstalk and insertion losses under realistic conditions of operation. Switching speed is expected in the ms-range, typical of nematic materials, which is sufficient for light reconfiguration and routing applications. As in other LC-based photonic components, power consumption of the device can be estimated as $P \simeq C f V_{\textrm{LC}}^2$, where $C$ is the device capacitance and $f=1\div 10$ KHz the LC-switching frequency, yielding a value in the deep sub-{\textmu}W regime. Such components are envisaged as low-power tunable light routing elements in integrated photonic architectures for optical inter-chip interconnects.

This work was supported by the Marie-Curie Intra-European Fellowship ALLOPLASM (FP7-PEOPLE-$2010$-IEF-$273528$), within the 7th European Community Framework Programme.

\bibliographystyle{nature}
\bibliography{ReferenceDatabase}

\end{document}